\documentclass[aps,preprint,nofootinbib]{revtex4}
\usepackage{amsmath}
\usepackage{amssymb}
\begin{document}
\author{R. N. Ghalati}
\email{rnowbakh@uwo.ca}
\affiliation{Department of Applied Mathematics,
University of Western Ontario, London, N6A~5B7 Canada}
\author{D. G. C. McKeon}
\email{dgmckeo2@uwo.ca}
\affiliation{Department of Applied Mathematics,
University of Western Ontario, London, N6A~5B7 Canada}
\title{\begin{flushright}
 {\footnotesize UWO\,-TH-\,07/17}
\end{flushright}A Canonical Analysis of the First Order Einstein-Hilbert Action}
\date{\today}

\begin{abstract}
The Dirac constraint formalism is applied to the $d\,(d>2)$ dimensional Einstein-Hilbert action when written in first order form,
 using the metric density and affine connection as independent fields. Field equations not involving time derivatives are not 
used to eliminate fields. Primary, secondary and tertiary constraints arise, leaving $d(d-3)$ degrees of freedom in phase space. 
The Poisson Bracket algebra of these constraints is given.
\end{abstract}
\maketitle
A full understanding of the Einstein-Hilbert action which could possibly lead to a consistent quantization of gravity is 
contingent upon understanding its canonical structure. Any canonical analysis depends on the choice of geometrical 
quantities that are taken to be the fundamental fields. Here we follow refs.\ [1-7] and work with the metric and affine 
connection; in particular we treat them as being independent as in refs.\ [3,7]. We differ from these papers however, 
as we do not use all the equations of motion which do not contain time derivatives to eliminate any fields. This is because 
some of these equations turn out to be secondary first class constraints that lead to further tertiary first class 
constraints when using Dirac's constraint formalism [8,9]\,. The resulting Poisson Bracket (PB) algebra of these 
constraints will possibly lead to an unsuspected symmetry in the action [11,12]\,.

The Einstein-Hilbert action $S_d=\int d^dx \sqrt{-g} R$ when $d>2$ can be written in first order form 

\begin{equation}\label{1}
S_d=\int d^d x \,h^{\mu \nu}\,(G^\lambda_{\mu \nu,\lambda}+\frac{1}{d-1} G^\lambda_{\lambda \mu} 
G^\sigma_{\sigma \nu}-G^\lambda_{\sigma \mu}
G^\sigma_{\lambda \nu})
\end{equation}
\vspace{.1cm}

\noindent where $h^{\mu\nu}=\sqrt{-g}g^{\mu\nu}$ and $G^\lambda_{\mu \nu} = \Gamma^\lambda_{\mu \nu}-\frac{1}{2}(\delta^\lambda_\nu \Gamma^\sigma_{\mu \sigma}+\delta^\lambda_\mu \Gamma^\sigma_{\nu \sigma})$\,.
If now $h=h^{00}$, $h^i=h^{0i}$, $\omega=-G^0_{00}+\frac{h^ih^j}{h^2}\,G^0_{ij}$, $\omega_i=-2G^0_{0i}-2\frac{h^j}{h}\,G^0_{ij}$, $\omega_{ij}=G^0_{ij}$, $\xi^i=-G^i_{00}$, $t=-2G^i_{i0}$, $\bar \zeta^k_l=-2G^k_{l0}+\frac{2}{d-1}\,\delta^k_lG^i_{i0}-\frac{2}{h}\,(G^k_{lm}-\frac{1}{d-1}\,\delta^k_lG^i_{im})h^m$, $\xi^i_{jk}=-G^i_{jk}$ and $H^{ij}=\frac{h^ih^j}{h}-h^{ij}$, then $S_d$ can be written as 

\begin{eqnarray}\label{2}
S_d &=& \int\,d^d\,x \bigg[\,\omega\, h_{,0}+\omega_i \,h^i_{,0}+\omega_{ij} \,H^{ij}_{\,\,,0}\\ \nonumber
    &+& \,\frac{2-d}{d-1}\bigg(h\,(\omega+\frac{1}{2}\,\frac{h^i\omega_i}{h})^2 - \frac{1}{4} H^{ij}
        (\omega_i+\frac{2\omega_{im}\,h^m}{h})(\omega_j+\frac{2\omega_{jn}\,h^n}{h})\bigg)\\ \nonumber
    &+& \xi^i \,\chi_i+\frac{t}{d-1}\,\chi+\bar \zeta^i_j\,\,\lambda^j_i+ \xi^i_{jk}\,\sigma^{jk}_i- 
        \frac{h}{4}\,\bar \zeta^i_j\,\bar\zeta^j_i+H^{ij}\, \left(\xi^k_{li}\,\xi^l_{kj}-\frac{1}{d-1} \,\xi^k_{ki}\, \xi^l_{lj}\right)
\bigg]\,,
\end{eqnarray}

\noindent where 

\begin{eqnarray}\label{3-6}
\chi&=&h^j_{,j}+h\,\omega-H^{jk}\,\omega_{jk}\,,\\ 
\chi_i&=&h_{,i}-h\,\omega_i\,,\\
\lambda^j_i&=&h^j_{,i}-\frac{1}{2}\,h^j\,\omega_i-H^{jk}\,\omega_{ik}\,,\\
\sigma^{jk}_i &=& -H^{jk}_{\,\,\,,i}+\frac{1}{h}(h^j\,H^{kl}+h^k\,H^{jl})\,\omega_{il}-\frac{1}{d-1}(\delta^j_i\,H^{kl}+
\delta^k_i\,H^{jl})(\frac{1}{2}\,\omega_l+\omega_{lm}\,\frac{h^m}{h})\,\\ \nonumber
&-& \frac{1}{h^2}\,h^jh^k\,\chi_i+\frac{1}{(d-1)h}\,(\,\delta^j_i\,h^k+\delta^k_i\,h^j\,)\,\chi\,.
\end{eqnarray}

The momenta associated with $\omega$, $\omega_i$ and $\omega_{ij}$ are all zero while the momenta associated with $h$, $h^i$, $H^{ij}$ 
are $\omega$, $\omega_i$ and $\omega_{ij}$\, respectively; this gives $d(d+1)$ primary second class constraints. As the momenta associated 
with $t$ and $\xi^i$ also vanish there are a further set of $d$ primary first class constraints. Reading the canonical Hamiltonian 
$H$ off of eq.\ (2), we find that the requirement that these first class constraints have a vanishing PB with $H$ leads to the secondary constraints

\begin{equation}\label{7}
\chi=\chi_i=0\,.
\end{equation} 

These have the PB algebra 

\begin{equation}\label{8}
\big\{\chi_i\,,\chi\big\}=\chi_i\,\,\,\,\,\,\,\,\,\,\,\,\,\,\,\,\,\,\,\,\,\,\,\,\,\,\,\,\,\,
\big\{\chi,\chi\big\}=0=\big\{\chi_i\,,\chi_j\big\}\,.
\end{equation}

The momenta associated with $\xi^i_{jk}$ and $\bar \zeta^i_j$ also vanish; these constitute primary constraints which lead to further secondary constraints which are the equations of motion of $\xi^i_{jk}$ and $\bar \zeta^i_j$; in total this gives $d(d^2-3)$ second class constraints. These equations of motion can be used to now eliminate $\xi^i_{jk}$ and $\bar \zeta^i_j$ in $H$. (We do not use eq.\ (7) to eliminate fields from $H$ as it will be shown that these are first class constraints which lead to tertiary first class constraints.)

If the Hamiltonian is to have a vanishing PB with $\chi$, then there is a further tertiary constraint

\begin{equation}\label{9}
\bar \tau=H+\delta^i_{,i}
\end{equation}

\noindent with $\delta^i=H^{ij}_{\,\,,j}-\frac{1}{h}(h^ih^j),_{j}-2h^i\omega+H^{ij}(\omega_j+\frac{2\omega_{jm}h^m}{h})$\,; similarly, $\chi_i$ leads to the tertiary constraint

\begin{equation}\label{10}
\tau_i=h(\frac{1}{h}H^{pq}\omega_{pq})_{,i}+H^{pq}\omega_{pq,i}-2(H^{pq}\omega_{qi})_{,p}\,.
\end{equation}

By adding a linear combination of constraints to $\bar \tau$\,, it can be reduced to

\begin{eqnarray}\label{11}
\tau &=& -H^{ij}_{,ij}-(H^{ij}\omega_j)_{,i}-\frac{d-3}{4(d-2)}\,H^{ij}\omega_i\omega_j+\frac{1}{2(d-2)}\,H_{kl}H^{kl}_{\,\,,i}H^{ij}\omega_j\\ \nonumber
&-&\frac{1}{h}H^{ik}H^{jl}(\,\omega_{jk}\,\omega_{il}-\omega_{ik}\,\omega_{jl})+\frac{1}{2}H^{jk}_{\,\,,i}H_{jl}H^{il}_{,k}
+\frac{1}{4}H^{ij}H_{kl,i}H^{kl}_{\,\,,j}\\ \nonumber
&+& \frac{1}{4(d-2)}H^{ij}H_{kl}H^{kl}_{\,\,,i}H_{mn}H^{mn}_{\,\,\,\,\,,j}\,.
\end{eqnarray}

The tertiary constraints have the PB

\begin{equation}\label{12}
\big\{\chi\,,\tau_i\big\}=0=\big\{\chi_i\,,\tau\big\}=\big\{\chi_i\,,\tau_j\big\}\,\,,\,\,\big\{\chi\,,\,\tau\big\}=\tau\,,
\end{equation}

as well as the non-local PB

\begin{eqnarray}\label{13}
\int dx\,dy\,&f(x)& \{\tau_i(x),\tau_j(x)\}\,\,g(y) 
= \int dx \left(\,gf_{,j}\,\tau_i-fg_{,i}\,\tau_j\right)\,,
\end{eqnarray}
\begin{eqnarray}\label{14}
\int dx\,dy\, &f(x)& \big\{\tau(x),\tau(x)\big\}\,\,g(y) \\ \nonumber
&=&\int dx \bigg[\left(gf_{,i}-fg_{,i}\right)\frac{H^{ij}}{h^2}\left(h\tau_j-H^{mn}\omega_{mn}\chi_j+
2H^{mn}\omega_{mj}\chi_n\right)\,\bigg]\,,
\end{eqnarray}
\begin{eqnarray}\label{15}
\int dx\,dy\,&f(x)& \big\{\tau_i(x),\tau(x)\big\}\,g(y)\\ \nonumber
&=& \int dx \bigg[g \frac{{(fh)}_{,i}}{h}\,\tau-fg_{,i}\tau-\frac{d-3}{2(d-2)}fgH^{kl}\omega_k\left(\frac{\chi_l}{h}\right)_{,i}\\ \nonumber
&-&\frac{d-3}{2(d-2)}gf_{,k}H^{kl}\,\omega_l\left(\frac{\chi_i}{h}\right)+f_{,k}\,g_{,l}\,H^{kl}\left(\frac{\chi_i}{h}\right)\\ \nonumber
&+& f g_{,k} H^{kl}\left(\frac{\chi_l}{h}\right)_{,i}+\frac{1}{2(d-2)}gf_{,m}H^{mn}H_{kl}H^{kl}_{\,\,,n}\left(\frac{\chi_i}{h}\right)\\ \nonumber
&+& \frac{1}{2(d-2)}fgH^{mn}H_{kl}H^{kl}_{\,\,,m}\left(\frac{\chi_n}{h}\right)_{,i}\bigg]\,,
\end{eqnarray}

\noindent where $f$ and $g$ are test functions.

Eqns.\ (8,12-15) show that $\left(\,\chi\,,\,\chi_i\,,\,\tau\, ,\, \tau_i\,\right)$ are all first class constraints. By eq.\ (9) we see that since $H$ and $\tau$ differ by a total divergence and a linear combination of constraints, no further constraints are generated by the consistency condition that all constraints have a weakly vanishing time derivative.

There are now $(d^3+d^2-2d)$ second class constraints, $3d$ first class constraints and $3d$ gauge conditions, placing $d(d^2+d+4)$ constraints on the $d(d+1)^2$ phase space variables, leaving $d(d-3)$ independent degrees of freedom. The expectations for the constraint structure of $S_d$ outlined in ref. [12] are thus realized.

It may be possible that functions of $\left(\,\chi\,,\,\chi_i\,,\,\tau\, ,\, \tau_i\,\right)$ would simplify the algebra of PBs of the first class constraints; this would facilitate determination of the gauge invariance of $S_d$ using the formalism of refs.\ [10,11]. The gauge structure for $S_2$ is discussed in ref.\ [12]\,. A detailed description of the calculations presented here will be forthcoming [13]\,.

\vspace{1cm}
\noindent
{\bf Acknowledgments}
\vspace{.1cm}

We would especially like to thank S. Kuzmin and N. Kiriushcheva for numerous fruitful discussions. F. T. Brandt and T. N. Sherry assisted in this work. R. Macleod had a helpful suggestion.


\end{document}